\documentclass[]{aa} 
\usepackage{graphicx}   
\usepackage{natbib}   
\usepackage{color}   
\bibpunct{(}{)}{;}{a}{}{,} 

\newcommand{\CII}{[\ion{C}{ii}]}   
   
\newcommand{\CI}{[\ion{C}{i}]}

\newcommand{\HII}{\ion{H}{ii}}   
\newcommand{\HI}{\ion{H}{i}}   
   
\newcommand{\NII}{[\ion{N}{ii}]}   
\newcommand{\lsun}{L$_{\odot}$}   
\newcommand{\msun}{M$_{\odot}$}   
\newcommand{\msol}{M$_{\odot}$}   

\def\fh{\hbox{$.\!\!^{\rm h}$}}
\def\fm{\hbox{$.\!\!^{\rm m}$}}
\def\fs{\hbox{$.\!\!^{\rm s}$}}
\def\fdg{\hbox{$.\!\!^\circ$}}
\def\farcm{\hbox{$.\mkern-4mu^\prime$}}
\def\farcs{\hbox{$.\!\!^{\prime\prime}$}}
\begin{document}   
%
   
\title{Clumpy photon-dominated regions in Carina}
\subtitle{I. \CI\ and mid-$J$ CO lines in two $4'\times4'$ fields.}
     
   \author{   
     C.\,Kramer\inst{1} \and
     M.\,Cubick\inst{1} \and
     M.\,R\"ollig\inst{3} \and
     K.\,Sun\inst{1} \and
     Y.\,Yonekura\inst{2} \and
     M.\,Aravena\inst{3} \and
     F.\,Bensch\inst{3} \and
     F.\,Bertoldi\inst{3} \and
     L.\,Bronfman\inst{4} \and
     M.\,Fujishita\inst{5} \and
     Y.\,Fukui\inst{5} \and
     U.U.\,Graf\inst{1} \and
     M.\,Hitschfeld\inst{1} \and
     N.\,Honingh\inst{1} \and
     S.\,Ito\inst{5} \and
     H.\,Jakob\inst{1} \and
     K.\,Jacobs\inst{1} \and
     U.\,Klein\inst{3} \and
     B.-C.\,Koo\inst{6} \and
     J.\,May\inst{4} \and
     M.\,Miller\inst{1} \and
     Y.\,Miyamoto\inst{5} \and 
     N.\,Mizuno\inst{5} \and
     T.\,Onishi\inst{5} \and
     Y.-S.\,Park\inst{6} \and
     J.L.\,Pineda\inst{3} \and
     D.\,Rabanus\inst{1} \and
     H.\,Sasago\inst{5} \and
     R.\,Schieder\inst{1} \and
     R.\,Simon\inst{1} \and
     J.\,Stutzki\inst{1} \and
     N.\,Volgenau\inst{1} \and
     H.\,Yamamoto\inst{5}
          }   


   \institute{   
     KOSMA, I. Physikalisches Institut, Universit\"at zu K\"oln,   
     Z\"ulpicher Stra\ss{}e 77, D-50937 K\"oln, Germany    
     \and
     Department of Physical Science, Osaka Prefecture University,
     Osaka 599-8531, Japan
     \and
     Argelander-Institut f\"ur Astronomie,  Auf dem H\"ugel 71,
     D-53121 Bonn, Germany
     \and
     Departamento de Astronom\'{i}a, Universidad de Chile, Casilla 36-D, Santiago, Chile
     \and
     Department of Astrophysics, Nagoya University, Chikusa-ku, Nagoya 464-8602, Japan
     \and
     Seoul National University, Seoul 151-742, Korea
   }   
   
   \offprints{C.\,Kramer, \email{kramer@ph1.uni-koeln.de}}   
   \date{Received / Accepted }   
      
   \abstract
   {The Carina region is an excellent astrophysical laboratory for
     studying the feedback mechanisms of newly born, very massive
     stars within their natal giant molecular clouds (GMCs) at only
     2.35~kpc distance.}
   {We use a clumpy PDR model to analyse the observed intensities of
     atomic carbon and CO and to derive the excitation conditions of the
     gas.}
   {The NANTEN2-4m submillimeter telescope was used to map the \CI\
     $^3P_1-^3P_0$, $^3P_2-^3P_1$ and CO 4--3, 7--6 lines in two
     $4'\times4'$ regions of Carina where molecular material
     interfaces with radiation from the massive star clusters. One
     region is the northern molecular cloud near the compact OB
     cluster Tr\,14, and the second region is in the molecular cloud
     south of $\eta$Car and Tr\,16. These data were combined with
     $^{13}$CO SEST spectra, HIRES/IRAS $60\,\mu$m and $100\,\mu$m
     maps of the FIR continuum, and maps of 8$\,\mu$m IRAC/Spitzer and
     MSX emission.}
   {We used the HIRES far-infrared dust data to create a map of the FUV
     field heating the gas. The northern region shows an FUV field of a
     few $10^3$ in Draine units while the field of the southern region
     is about a factor 10 weaker. While the IRAC 8$\,\mu$m emission
     lights up at the edges of the molecular clouds, CO and also \CI\
     appear to trace the H$_2$ gas column density.  The northern
     region shows a complex velocity and spatial structure, while the
     southern region shows an edge-on PDR with a single Gaussian
     velocity component.  We constructed models consisting of an
     ensemble of small spherically symmetric PDR clumps within the
     $38''$ beam (0.43~pc),
     which follow canonical power-law mass and mass-size
     distributions. We find that an average local clump density of
     $2\,10^5$\,cm$^{-3}$ is needed to reproduce the observed line
     emission at two selected interface positions. }
   {Stationary, clumpy PDR models reproduce the observed cooling lines
     of atomic carbon and CO at two positions in the Carina Nebula. }
%
  
   \authorrunning{Kramer et  al.} 
   \maketitle   

\section{Introduction} 

\subsection{The importance of atomic carbon}

The fine structure lines of atomic carbon are strong coolants of the
interstellar medium (ISM) throughout the universe: in Galactic clouds
\citep[e.g.][]{jakob2007}, in the Milky Way as a whole
\citep{fixsen1999}, as well as in external galaxies
\citep{bayet2006,kramer2005} up to high redshifts
\citep[][]{weiss2003}. Large-scale observations of Galactic clouds
show \CI\ emission coincident with that of CO.
This holds for giant molecular clouds (GMCs) \citep{sakai2006}, as well
as for rather quiescent dark clouds \citep{tatematsu1999}. It has
therefore been suggested that \CI\ is a more reliable tracer of the
molecular cloud masses in external galaxies than CO
\citep{papadopoulos2004}. However, all steady-state models of
photon-dominated regions (PDRs) predict that atomic carbon is formed
in a surface layer by recombination of C$^+$ and dissociation of CO.
This appears to be in contradiction with atomic carbon emission which
is observed even at large distances from UV sources. The dynamics of
the ISM may play a decisive role, allowing to explain the distribution
of \CI\ with time dependent chemical models
\citep{oka2004,stoerzer1997}.

However, stationary, but clumpy PDR models have been successful in
explaining e.g.  extended \CI\ $^3P_1-^3P_0$ and $^{13}$CO\,$2-1$
emission in S\,140 \citep{spaans1996,spaans1997}.  Inside a clumpy
PDR, the UV radiation can penetrate much deeper than it would be
possible in a homogeneous layer \citep{boisse1990}.
\citet{meixner1993} showed that the line intensities of all lines are
enhanced compared to a homogeneous model with the same average
density.
  \cite{juvela2001} presented a model to apply Monte Carlo radiative
  transfer calculations to the results of magnetohydrodynamic
  calculations to account for a clumpy and turbulent structure. They
  also found that local density condensations lead to an increased
  cooling efficiency and reduced photon trapping. 
  Another approach is to describe both the turbulent velocity field
  and the density structure by their statistical properties.
  \cite{hegmann2007} calculate CO cooling rates based on a stochastic
  radiative transfer model, which accounts for density and velocity
  fluctuations with a finite correlation length.

  Here, we study the distribution of atomic carbon and
  warm CO in the Carina star forming region and test clumpy models
  using the KOSMA-$\tau$ PDR model \citep{stoerzer1996,roellig2007} to
  derive the excitation conditions of the gas. Spaans et al. model the
  inhomogeneous structure with a Monte-Carlo code, assuming a fixed
  clump/interclump density ratio of 10. In contrast, we assume that
  all emission analyzed stems from the clumps. The influence of the
  interclump medium or a halo leading to pre-shielding of CO has been
  discussed by \citet{bensch2006} and \citet{bensch2003}, who also
  used KOSMA-$\tau$, but is not considered here.  To describe the
  clumpy structure, we assume that the clumps are distributed
  following the canonical clump mass and mass-size distributions which
  have been found in molecular clouds.

\begin{figure*}[ht]   
  \centering   
   \includegraphics[angle=-90,width=18cm]{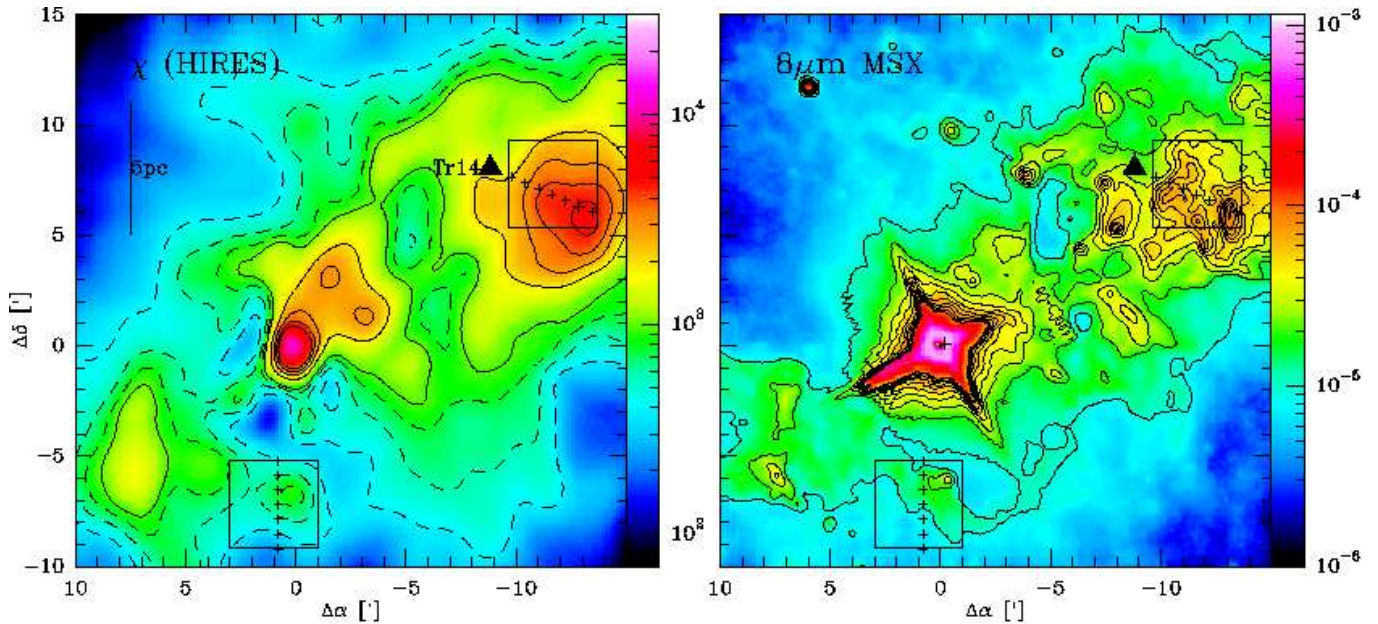}
  \caption{Overview of the Carina clouds near $\eta$Car lying at 0/0
    10:45:03.59 -59:41:04.3 (J2000) at the center of the Tr\,16
    cluster.  {\bf Left:} Estimate of the FUV flux $\chi$ in Draine units  at a
    resolution of $\sim1'$ and derived from HIRES/IRAS. Dashed
    contours range from 400 to 800 in steps of 200.
      Solid contours range from $10^3$ to $5\,10^3$ in steps of
      $10^3$.
%
    {\bf Right:} MSX emission at 8\,$\mu$m at a resolution of
    $\sim20''$.  Contours are 1 to 10 in steps of $1~10^{-5}$
    Wm$^{-2}$sr$^{-1}$ \citep[cf.  Figs.\,7, 11 in][]{rathborne2002}.
    Note the diffraction artefacts near $\eta$Car.  The center
    position of the OB cluster Tr\,14 is marked by a triangle.  The
    two $4'\times4'$ regions mapped with NANTEN2 are delineated by
    boxes.  Crosses in the left panel mark two cuts discussed in the text.  }
\label{fig-overview}   
\end{figure*}   

\begin{center}   
\begin{table*}[ht]   
  \caption[]{\label{tab-carn-wco} {\small Physical parameters derived
      from the observed integrated intensities $I$ in K\,kms$^{-1}$ and
      intensity ratios $R^{74}_{\rm CO}$=$I$(CO 7--6)/$I$(CO 4--3),
      $R^{21}_{\rm C}$=$I$(\CI 2--1)/$I$(\CI 1--0), $R^{14}_{\rm
        CCO}$=$I$(\CI 1--0)/$I$(CO 4--3) from Gaussian fits to the
      line profiles at two interface positions.
      %
      The calibration error of the
      integrated intensities is estimated to be 15\%.
}}
\begin{tabular}{lllllllllrrrrrr}   
\hline \hline   
  $\Delta\alpha/\Delta\delta$ & $^{13}$CO & \multicolumn{2}{c}{CO} & \multicolumn{2}{c}{\CI} & 
   $R^{74}_{\rm CO}$ & $R^{21}_{\rm C}$ & $R^{14}_{\rm CCO}$ &
 $T_{\rm ex}$ & $N$(CO) & $N$(C) & $N$(H$_2$) & M & C/CO \\
  $(',')$ & 2--1 & 4--3 & 7--6 & 1--0 & 2--1 & & & & K & 
  & & & \msol \\
\hline
$-11.7/6.8$  & 47 & 227 & 86 & 37 & 27 & 0.38 & 0.72 & 0.16 
   &  50 &  23$^{30}_{18}$ &   4.9$^{5.8}_{4.1}$ &  28$^{35}_{21}$ &  100 
   & 0.21$^{0.32}_{0.13}$ \\
$+0.8/-7.2$    & 40  & 90 & 24 & 31 & 19 & 0.27 & 0.6 & 0.34 
 &  30 &  15$^{18}_{22}$ &   3.8$^{4.5}_{3.3}$ &  18$^{22}_{14}$ &  63 
 & 0.25$^{0.38}_{0.18}$ \\
\noalign{\smallskip} \hline \noalign{\smallskip}   
\end{tabular}   
\end{table*}   
\end{center}   

\subsection{The Carina nebula}

  The Carina nebula is a very prominent southern
  star-forming region at a distance of 2.35~kpc
    \citep{smith2006b}.  This giant molecular cloud was first
  observed by \citet{grabelsky1988} as part of the Columbia CO 1--0
  survey of the Milky Way. They derived a total mass of
  $6.7\,10^5$\,\msun\ and an effective radius of 66\,pc.  In total 65
  O-type stars, including 6 of the 11 O3-type stars known in the Milky
  Way \citep{smith2006}, produce a strong UV field and strong stellar
  winds which interact vigorously with the material in the surrounding
  molecular clouds.  This high concentration of the earliest-type
  stars is unique in the Galaxy and may serve as a template for more
  extreme but more distant regions containing stellar super clusters
  like the central regions of NGC\,253 or the Antennae
  \citep{bayet2006,schulz2007}.  The Carina nebula extends more than 4
  square degrees on the sky.  It exhibits a peculiar morphological
  structure on all scales.  Massive stars are being born in several
  regions within molecular condensations, located mainly to the
  south-east and north-west of the cluster Trumpler\,16, which is
  centered at $\eta$Car, a luminous blue variable star.  $\eta$Car is
  embedded in the Homunculus nebula \citep[e.g.][]{smith2006} and its
  luminosity has changed drastically and repeatedly in the past
  centuries, leading to strong variations of the FUV field impinging
  on the surrounding clouds. Its current luminosity of
  $5~10^6$\,\lsun\ makes it one of the brightest celestial infrared
  objects \citep{cox1995}. In the northern part of the Carina nebula,
  the stellar population is dominated by the cluster Tr\,14.
Recently, \citet{oberst2006} reported the detection of the \NII\ line
at $205\,\mu$m at the radio peak of the Carina II \HII\ region using
SPIFI at AST/RO. They compared their data with \CII\ data
\citep{mizutani2004,mizutani2002} and concluded that about a third of
the \CII\ emission stems from a low-density ionized medium.
%
For a recent overview of the Carina region, see \citet{smith_brooks2007}.




%

Carina was mapped at $\sim3'$ resolution in the CO 4--3 \newline and \CI\
$^3P_1-^3P_0$ lines by \citet{zhang2001} using the AST/RO 1.7~m
telescope. These AST/RO data show large-scale \CI\ emission,
coextensive with CO 4--3.


Here, we have used the 4~m NANTEN2 telescope to map $^{12}$CO in the
rotational transitions $J=$4--3, 7--6, and \CI\ in the fine structure
transitions $^3P_1-^3P_0$, and \newline $^3P_2-^3P_1$ (henceforth 1--0
and 2--1) at sub-arcminute resolutions in two $4'\times4'$ regions of
Carina (Fig.\,\ref{fig-overview}). 

One region to the north of $\eta$Car covers the cloud interface with
Tr\,14.  The second region covers another cloud interface to the south
of $\eta$Car and Tr\,16.


The northern molecular cloud to the west of Tr\,14
(Fig\,\ref{fig-overview}) contains several sites of massive star
formation.  Diffuse PAH $3.29\,\mu$m emission has been found by
\citet{rathborne2002} to trace the sharp edge of the cloud.  Peak
intensities at the edge towards the Car\,I \HII\ region are almost 10
times higher than in the other regions of Carina.  Car\,I-E is
interacting with the front face of the GMC and creating a PDR seen
edge-on \citep{brooks2003,brooks2001}.  Car\,I-W manifests a second
PDR seen face-on (cf. Fig.\,\ref{fig-carina-n-spitzer}).

The CO 2--1 map of \citet{cox1995} shows a sharp rim of gas
delineating the southern molecular cloud, located south of $\eta$Car
and the Tr\,16 cluster.  Several IRAS sources are located along the
edge. A prominent peak in PAH 3.29~$\mu$m emission is the ultra
compact \HII\ region IRAS\,$10430-5931$ \citep{rathborne2002}.

\section{NANTEN2 observations} 

We used the new NANTEN2 4m telescope situated at 4865\,m altitude at
Pampa La Bola in northern Chile to map two regions in Carina in CO
4--3 and 7--6, and in the two fine structure transitions of atomic
carbon.  Observations were conducted in 2006 between September 21 and
October 6 with a dual-channel 460/810 GHz receiver installed for
verifying the telescope submillimeter performance.  Double-sideband
(DSB) receiver temperatures were $\sim250$\,K in the lower channel and
$\sim750$\,K in the upper one. The intermediate frequencies (IF) are
4\,GHz and 1.5\,GHz, respectively. The latter IF allows simultaneous
observations of the CO 7--6 line in the lower and the \CI\ 2--1 line
in the upper sideband. These two lines are observed simultaneously
with one of the lines in the 460\,GHz channel.  As backends, we used
two acousto-optical spectrometers (AOS) with a bandwidth of 1\,GHz and
a channel resolution of 0.37\,kms$^{-1}$ at 460\,GHz and
0.21\,kms$^{-1}$ at 806\,GHz.

The pointing was checked regularly on Jupiter, IRC$+10216$, and IRc2
in Orion\,A. The applied corrections were always smaller than $20''$,
and usually less than $10''$.  The atmospheric transmission was
derived by measuring the atmospheric emission at the reference
position. Carina was observed between $\sim32^\circ$ and
$\sim54^\circ$ elevation. Spectra of the two frequency bands were
calibrated separately, and sideband imbalances were corrected using
the atmospheric model {\tt atm} \citep{pardo2001}.  We used a
reference position at 10\fh50\fm09\fs3 $-59$\fdg18\farcm33\farcs7
(J2000), $\sim45'$ to the north-east of $\eta$Car,
which was observed to be free of emission by \citet{yonekura2005}.
Observations were conducted on-the-fly (OTF), scanning in right
ascension at a speed of $2.5''$/sec and with a sampling interval of
$10''$.  \CI\ emission of the northern field was also observed
scanning in declination. The reference position was observed at the
beginning of each OTF scanning line, i.e.  every 1.6\,minutes in time.
To complete a single scan of the $4'\times4'$ map area took
$\sim1\,$hour of total observing time.  ON-source integration times
per position on a $10''$ grid range between 24\,sec (for CO 4-3 \& 7-6
in the northern field) and 4\,sec. The total observing time was
10\,hours.

%

The half power beamwidths (HPBWs) and main beam efficiencies $(B_{\rm
  eff})$ were determined from continuum cross scans on Jupiter
\citep{simon2007}. For the latter, we linearly interpolated the
Jovian brightness temperatures listed by \citet{griffin1986} to the
observed frequencies.
%
The HPBWs deconvolved from the observed full widths at half maximum
(FWHM), are 38\farcs0 and 26\farcs5, in the lower and upper receiver
band, respectively.  Beam efficiencies are 50\% and 45\%,
respectively.  The forward efficiency $(F_{\rm eff})$ was derived from
skydips and was found to be 86\% for both frequency bands.  The raw
data taken at the telescope was recalibrated to the $T_{\rm A}^*$
antenna temperature scale and then multiplied by $F_{\rm eff}/B_{\rm
  eff}$ to scale to main beam temperatures.  All data presented here
are on the $T_{\rm mb}$ scale.  We fitted and subtracted 2nd order
polynomials from all spectra. In a few cases, we subtracted
higher-order polynomials.  

  During the test campaign in the second half of 2006,
  we were able to roughly measure the error beam amplitudes of the
  NANTEN2 antenna \citep{simon2007}.
  Continuum measurements of the solar and lunar edges allowed to
  measure the extent and strength of a first compact error beam at
  810\,GHz following the method described e.g. by \citet{greve1998}.
  We find a FWHM of $240''$ corresponding roughly to the average panel
  size of about 0.6\,m.  Only about 10--15\% of the power is detected
  within this error beam which we attribute to panel misalignment.
  The power detected within a beam of $\sim30'$ was derived from lunar
  continuum scans using a lunar Rayleigh-Jeans temperature of 350\,K
  for both frequency bands \citep{pardo2005,mangum1993}. Moon
  efficiencies are $\sim70\%$ at both frequencies.  
  In total, about 36\% of the total power is detected in at least two
  error beams at both frequencies.

  As the spatial source structure in each velocity
  channel is convolved by the antenna diagram during observations, the
  error beams lead to pickup of extended emission in Carina. Main beam
  temperatures presented here are somewhat too high. Line profiles may
  be slightly distorted \citep[cf.][]{bensch2001aa365-285}. However,
  as the error beams are not yet completely characterized and may
  moreover be somewhat time variable, we ignore their influence in the
  following. To first order, \CI\ and CO emission stems from the same
  region.  Line ratios will therefore be much less affected by the
  error beam compared to absolute intensities.  

\section{Complementary data}

\subsection{$^{13}$CO 2--1 data}

Brooks et al. obtained complementary maps of $^{13}$CO 2--1 at the
Swedish-ESO Submillimetre Telescope (SEST).  The telescope main beam
efficiency at 220\,GHz was 50\% and its HPBW $25''$.  The northern
$4'\times5'$ SEST map \citep{brooks2003} covers the northern NANTEN2
field. The southern map (Brooks, Schneider, priv.  comm.) covers the
southern NANTEN2 field. We convolved these data with a Gaussian kernel
to a resolution of $38''$ to allow a comparison with the NANTEN2
spectra taken along two cuts through the interface regions.

\subsection{Dust and PAHs in Carina}

Figure\,\ref{fig-overview} gives an overview of a $25'\times25'$
region surrounding $\eta$Car, showing the Keyhole region in the
center, and the northern and southern clouds.  The OB stars of Tr\,14
and Tr\,16 emit far-UV photons which are partly absorbed by the dust
grains of the surrounding molecular clouds. The heated dust cools in
the far-infrared via dust continuum emission. Gas heating is very
inefficient; only a few percent at most of the absorbed energy is
transferred to the gas \citep{hollenbach1999}, which cools via
far-infrared and submillimeter line emission. Since the FUV photons
play a decisive role in driving the cloud chemistry, determination of
the FUV field is important.

  To derive a map of the estimated FUV fluxes
  (Fig.\,\ref{fig-overview}), we obtained HIRES/IRAS 60 and
  100\,$\mu$m images at $\sim1'$ resolution from the IPAC data center.
  These images at enhanced resolution were created using a maximum
  correlation method \citep{aumann1990}. The filter properties of IRAS
  allow us to combine these two data sets to create a map of
  far-infrared intensities $I_{\rm FIR}$ between 42.5\,$\mu$m and
  122.5\,$\mu$m \citep{helou1988,nakagawa1998}.  
%
%
Assuming that the FUV energy absorbed by the grains is reradiated in
the far-infrared, we then estimate the FUV flux $\chi$ (6\,eV $<$
h$\nu < $13.6\,eV) impinging onto the cloud surfaces from the emergent
FIR intensities:
  $$\chi/\chi_0 = 4\pi~I_{\rm FIR}/({\rm erg\,s}^{-1}{\rm
    cm}^{-2}{\rm sr}^{-1})$$ with $\chi$ in units of
  $\chi_0=2.7\,10^{-3}$\,erg\,s$^{-1}$\,cm$^{-2}$
  \citep{draine1978,draine1996}.  Here, we assume that heating by
  photons with h$\nu<6$\,eV contributes a factor of $\sim2$
  \citep{tielens1985} and that the bolometric dust continuum intensity
  is a factor of $\sim2$ larger than $I_{\rm FIR}
  $\citep[][]{dale2001}. In this case, the two corrections cancel out.
  In case the clouds do not fill the beam, the derived estimate of the
  FUV field will only be a lower limit.  Obviously, this method fails
  and the FUV field cannot be derived at positions without dusty
  clouds.
%
%
%
%

    The northern $4'\times4'$ region mapped with NANTEN2 shows FUV
    field strengths varying between $2\,10^3$ in the north-east and
    $5\,10^3$ in the south-west. The FUV field in the southern region
    is a factor 10 weaker, varying between $4\,10^2$ in the outskirts
    and $8\,10^2$ near the center.

\begin{figure}[h]   
  \centering
  \includegraphics[angle=-90,width=8.5cm]{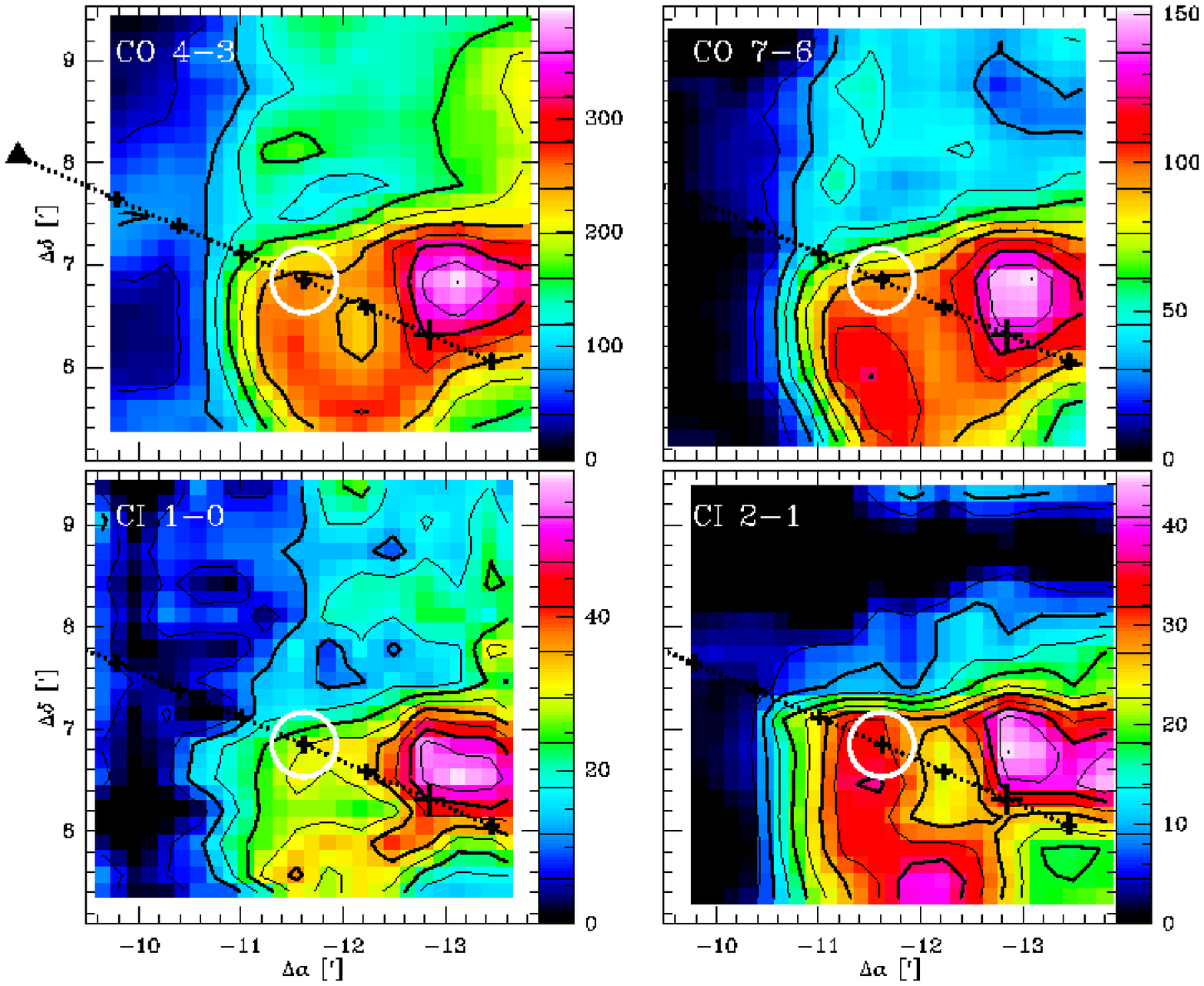}
  \caption{Northern $4'\times4'$ field: Velocity-integrated maps of CO
    4--3, 7--6, \CI\ 1--0, and 2--1 at a common angular resolution of
    $38''$ (0.43\,pc), integrated over the full
    velocity range of emission between $-32$ and $-5$\,kms$^{-1}$.
    Contours range between 10 and 90\% in steps of 10\% of the peak
    intensities, which are 398\,K\,kms$^{-1}$ for CO 4--3,
    153\,K\,kms$^{-1}$ for CO 7--6, 59\,K\,kms$^{-1}$ for \CI\ 1--0,
    and 48\,K\,kms$^{-1}$ for \CI\ 2--1.
The center position of the compact OB cluster Tr\,14 is marked by a
filled triangle outside the CO 4--3 map.
A dashed line and small crosses spaced by $40''$ mark a cut through
the peak of \CII\ emission \citep{brooks2003} (large cross) and
Tr\,14. The circle marks the resolution and the interface position
analyzed in Table\,\ref{tab-carn-wco}. }
\label{fig-carina-n-co43-ci10}   
\end{figure}

%
  \citet{brooks2003} estimate the FUV field at the
  position of peak \CII\ emission in the northern cloud
  (Figs.\,\ref{fig-overview},\ref{fig-carina-n-co43-ci10}) from the
  spectral types of the 13 brightest stars dominating the stellar
  content of Tr\,14. They derive $G_0=1.4\,10^4$ in units of the
  Habing field, i.e.  $\chi=8.2\,10^3$, a factor 1.6 larger than the
  peak flux in the northern region, as derived from HIRES.  However,
  the value derived by \citet{brooks2003} is a strict upper limit as
  it is assumed that the stars and cloud lie at the same distance to
  the observer and attenuation by dust can be neglected.  Moreover,
  the FUV field derived from the stars would be lowered by smearing to
  the resolution of the HIRES data. 


Figure\,\ref{fig-overview} shows contours of the FUV field together
with an $8\,\mu$m MSX image. The MSX image shows the emission from
very small grains and PAHs \citep{smith2006} excited by the FUV field.
Evidence that the PAH emission is stimulated by FUV absorption is
reflected by the good correlation between the FIR continuum and the
$8\,\mu$m emission.

\begin{figure}[h]   
  \centering   
  \includegraphics[angle=-90,width=7cm]{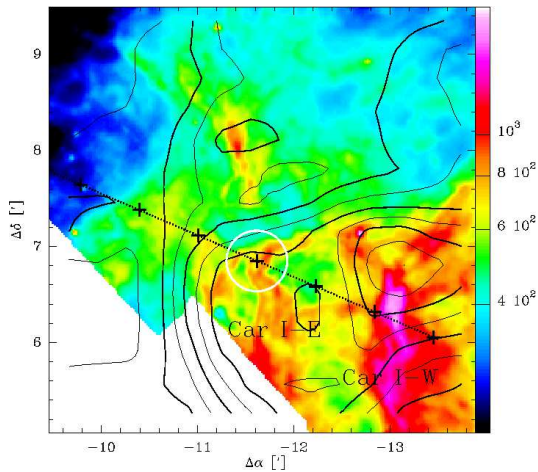}
  \caption{Northern field: Contours of integrated CO 4--3 intensities
    (cf.  Fig.\,\ref{fig-carina-n-co43-ci10}) overlayed on 8\,$\mu$m
    IRAC/Spitzer emission in units of MJy/sr. Car\,I-E and Car\,I-W
    are two \HII\ regions described e.g. by \citet{brooks2003} (cf.
    Fig.\,\ref{fig-carina-n-co43-ci10} for more details). }
%
\label{fig-carina-n-spitzer}   
\end{figure}   

\begin{figure}[h]   
  \centering
  \includegraphics[angle=-90,width=8.5cm]{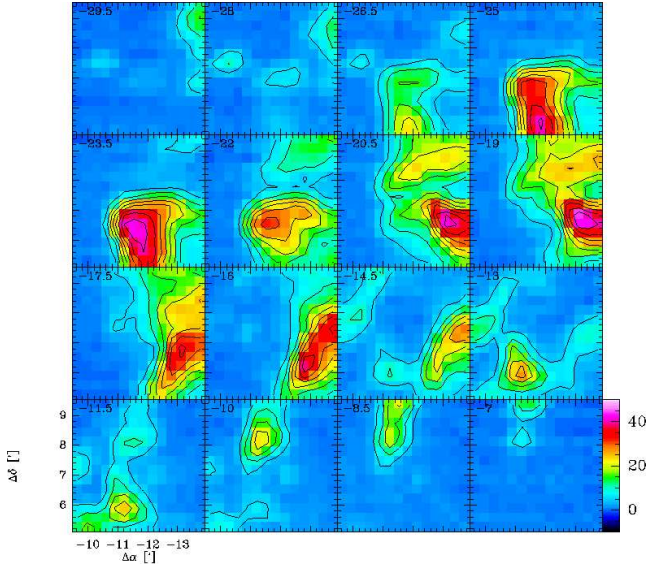}
\caption{Velocity structure of the northern field: CO 4--3 velocity
channels of 1.5\,kms$^{-1}$ width.  Contours range between 5 and
45\,K\,kms$^{-1}$ in steps of 5\,K\,kms$^{-1}$. }
\label{fig-carina-n-co43-channels}   
\end{figure}   

\begin{figure}[h]   
  \centering   
  \includegraphics[angle=-90,width=8.5cm]{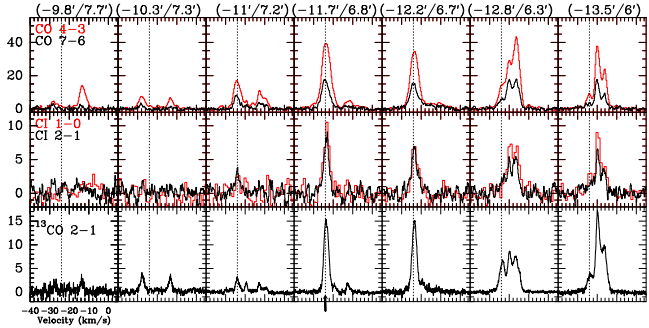}
  \caption{Spectra along a cut through the northern field connecting
    Tr\,14 and the \CII\ peak on a $40''$ grid. All data are at a
    common resolution of $38''$ and on the main beam scale.  The
    velocity range is $-40$ to $5$\,kms$^{-1}$.  The dashed vertical
    lines and the arrow mark the position of the $-24$\,kms$^{-1}$
    velocity component presented in Table\,\ref{tab-carn-wco}. }
%
%
\label{fig-carina-n-cut-spectra}   
\end{figure}   

\section{Results} 

\subsection{Northern region}

\subsubsection{Maps of integrated intensities}

\begin{figure}[h]   
  \centering \includegraphics[angle=-90,width=7cm]{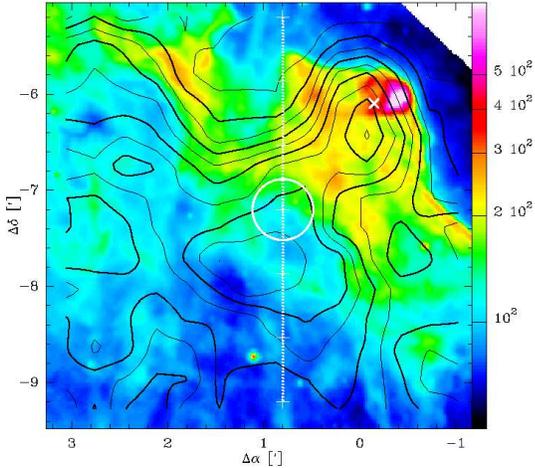}
\caption{Southern field: Contours of integrated CO 4--3 intensities
    (cf.  Fig.\,\ref{fig-carina-s-co43-ci10}) overlayed on 8\,$\mu$m
    IRAC/Spitzer emission in units of MJy/sr. }
%
\label{fig-carina-s-spitzer}   
\end{figure}

Figure\,\ref{fig-carina-n-spitzer} shows an overlay of integrated CO
4--3 intensities on an 8\,$\mu$m image taken with the Infrared Array
Camera (IRAC) on the Spitzer space telescope\footnote{These post-BCD
  data were retrieved from the Spitzer archive, reprojected, and
  merged to create Figs.\,\ref{fig-carina-n-spitzer} and
  \ref{fig-carina-s-spitzer} \citep[cf.][]{smith2005aas}.}.  The
8\,$\mu$m data reveal the complicated spatial structure of the region,
consisting of several filaments and shell-like structures. The bright
region in the south-west is the ionisation front Car\,I-W.
Approximately $\sim1'$ to the east lies Car\,I-E
\citep[cf.][]{rathborne2002,brooks2001}.
%

Maps of integrated CO 4--3, 7--6, \CI\ 1--0, and 2--1 emission
(Fig.\,\ref{fig-carina-n-co43-ci10}) show the boundary of the northern
molecular cloud (orientated north-south) towards the Tr\,14 cluster
and its \HII\ region.  For all four tracers, the integrated
intensities are co-extensive and peak in the south-west, near
$-13'/+6.5'$, i.e.  near the peak of \CII\ emission studied by
\citet{brooks2003}. The south-eastern interface region exhibits
increased CO 7--6 and \CI\ 2--1 line intensities, indicating an
increased amount of warm, dense gas. The morphology of the cloud
resembles a ring centered on a region with less strong emission near
$-12'/+$6\farcm5.  The northern half of the cloud shows weaker and
more diffuse emission.


%

The northern region exhibits a complex velocity structure between
$-32$ and $-5$\,kms$^{-1}$, which can be seen in the CO 4--3 velocity
channel maps (Fig.\,\ref{fig-carina-n-co43-channels}).  Note that weak
CO 4--3 emission is also detected east of the north-south interface,
near $-10'/+$7\farcm5, at velocities between about $-14.5$ and
$-10$\,kms$^{-1}$. The CO 4--3 emission also corresponds well with the
$^{13}$CO 2--1 SEST channel maps at $25''$ resolution shown in Fig.\,7
of \citet{brooks2003}. The good correspondance shows that the velocity
structure is largely due to individual clumps and not self-absorption
of optically thick $^{12}$CO.

\subsubsection{Spectra along a cut through the region}

The rich kinematical structure is further revealed by the spectra
along a cut, shown in Fig.\,\ref{fig-carina-n-cut-spectra}, from the
\CII\ emission peak through the cloud interface, towards Tr\,14.  This
cut is orientated the same as the cut presented in Figs.\,1 and 11 of
\citet{brooks2003}. The spectra of CO 4--3, 7--6, \CI\ 1--0, 2--1, and
$^{13}$CO 2--1 \citep{brooks2003} show several velocity components
with large velocity gradients.
At the interface, all tracers show a strong component at
$-24$\,kms$^{-1} $. Further west, this component shifts to lower
velocities and splits up into at least 3 components.
The line centroids of the different tracers match very well. 
The peak temperatures are 40\,K in CO 4--3 and 10\,K in \CI\ 1--0.
Line intensities of CO and \CI\ stay constant within a factor of 2 at
the positions within the cloud; they increase slightly at the
interface, before dropping rapidly towards the east. A weak component
at $-15$\,kms$^{-1}$ seen in CO and $^{13}$CO peaks to the east of the
interface at ($-$9\farcm8,7\farcm7).

To give one typical example, Table\,\ref{tab-carn-wco} shows the
integrated line intensities and derived quantities of the
$-24$\,kms$^{-1}$ component at the interface position, after Gaussian
fitting. 
$T_{\rm ex}$ is derived from the $^{12}$CO 4--3 peak line temperature
assuming optically thick emission and a beam-filling factor of 1.
Total CO and C column densities, given in $10^{17}$cm$^{-2}$, were
derived from $^{13}$CO 2--1 and \CI\ 1--0 assuming optically thin
emission at $T_{\rm ex}$, LTE, and a $^{12}$CO vs. $^{13}$CO abundance
ratio of 65.  Total H$_2$ column densities, given in
$10^{21}$cm$^{-2}$, and masses were derived from $N$(CO) assuming an
abundance ratio of $8.5\,10^{-5}$ \citep{frerking1982}.  C/CO is the
abundance ratio $N$(C)/$N$(CO).  Errors given for the column densities
were derived by varying $T_{\rm ex}$ and the integrated intensities by
$\pm15\%$.
%
While the $^{12}$CO 4--3 and 7--6 lines show FWHMs of 5.4 and
4.7\,kms$^{-1}$, the \CI\ 1--0 and 2--1 line widths are smaller, 3.9
and 3.1\,kms$^{-1}$.  $^{13}$CO 2--1 shows a still narrower line width
of 2.9\,kms$^{-1}$. We attribute this change of line widths to optical
depth effects.
%
%
The fitted line center velocities of all tracers agree within
0.5\,kms$^{-1}$. Line ratios of CO 7--6/4--3 and \CI\ 2--1/1--0 are
0.4 and 0.7, which are typical values for the entire cut. 

The peak CO 4--3 temperature of 40\,K translates, through the
Rayleigh-Jeans correction, into a lower limit of the gas kinetic
temperature of the emission zone of 50\,K. Optical depth effects or a
beam-filling factor of less than 1 would imply higher gas kinetic
temperatures.  \citet{brooks2003} derived a dust temperature of 50\,K
from the $60$ vs.  $100\,\mu$m IRAS flux ratio at the \CII\ position.

The \CI\ 2--1/1--0 line ratio is a sensitive function of the \CI\
excitation temperature.  In the optically thin limit and assuming LTE,
$T_{\rm ex}=38.3\,{\rm K}/\ln[2.11/R_{\rm CI}^{21}]$.  At the
interface position, we observed a ratio of $0.72\pm0.15$, assuming a
calibration error of the ratio of about 20\%. This translates into
$T_{\rm ex}$(\CI) of only 35\,K($^{43\,{\rm K}}_{29\,{\rm K}}$).

The $^{13}$CO 2--1 spectra \citep{brooks2003} peak at 15\,K. At the
interface, the \CI\ 1--0/$^{13}$CO 2--1 ratio is 0.8.


  To derive a first estimate of the total CO column
  densities at the two interface positions
  (Table\,\ref{tab-carn-wco}), we used the $^{13}$CO 2--1 integrated
  intensities assuming optically thin emission, LTE, and a $^{12}$CO
  vs. $^{13}$CO abundance ratio of 65.  The abundance ratio is in
  accordance with the average local ISM $^{12}$C/$^{13}$C ratio of
  $68\pm15$ recently found by \citet{milam2005}, which is valid also
  for Carina at about solar galacto centric distance. The inferred
  total optical extinction at the northern interface is $28\pm7$\,mag,
  using the canonical $N$(H$_2$)/A$_{\rm v}$ ratio of
  $9.36\,10^{20}$\,cm$^{-2}$\,mag$^{-1}$ \citep{bohlin1978}.  The C/CO
  abundance ratio is 0.21.

\begin{figure}[h]   
  \centering   
  \includegraphics[angle=-90,width=8.5cm]{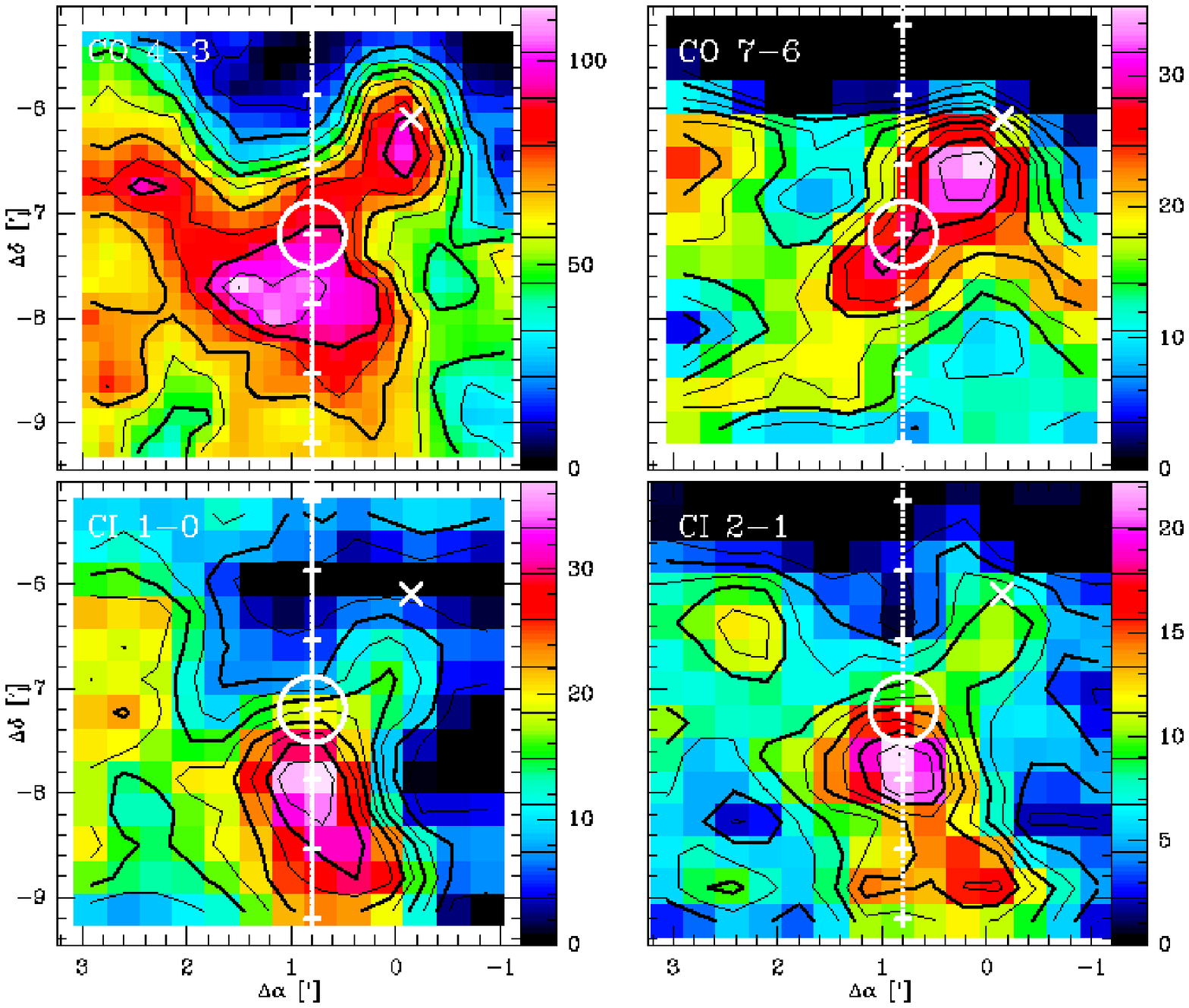}
  \caption{Southern $4'\times4'$ field: Velocity-integrated maps of CO
    4--3, 7--6, \CI\ 1--0, and 2--1 at a common angular resolution of
    $38''$, integrated over the full velocity range of emission
    between $-20$ and $-32$\,kms$^{-1}$.  Contours range between 10
    and 90\% in steps of 10\% of the peak intensities which are
    114\,K\,kms$^{-1}$ for CO 4--3, 35\,K\,kms$^{-1}$ for CO 7--6,
    37\,K\,kms$^{-1}$ for \CI\ 1--0, and 22\,K\,kms$^{-1}$ for \CI\
    2--1. A cross $\times$ marks the position of IRAS\,10430-5931 at
    10\fh45\fm02\fs4 $-59$\fdg47\farcm10\farcs0 (J2000).  Small
    crosses $+$ mark positions along a cut through the interface
    region.  The circle marks the resolution and the position analyzed
    in Table\,\ref{tab-carn-wco}. }
\label{fig-carina-s-co43-ci10}   
\end{figure} 

\begin{figure}[h]   
  \centering   
  \includegraphics[angle=-90,width=8.5cm]{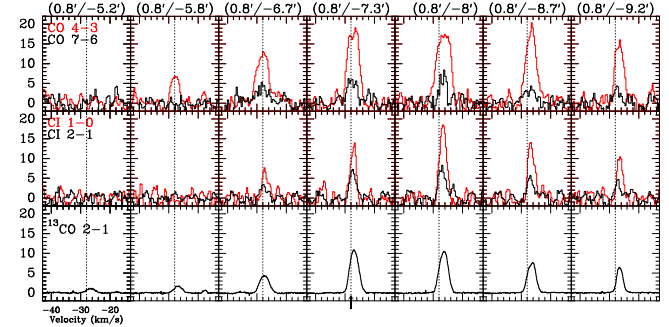}
  \caption{Spectra along a north-south orientated cut through the
    southern field at $\Delta\alpha=$0\farcm8 on a $40''$ grid. All
    data are at a common resolution of $38''$. The velocity range is
    $-43$ to $-13$\,kms$^{-1}$.  Vertical lines mark a velocity of
    $-28$\,kms$^{-1}$. The arrow marks the position analyzed in
    Table\,\ref{tab-carn-wco}.  
%
}
\label{fig-carina-s-co43-channels}   
\end{figure}   

\subsection{Southern region}

\subsubsection{Maps of integrated intensities}

Figure\,\ref{fig-carina-s-spitzer} shows an overlay of integrated CO
4--3 intensities on an 8\,$\mu$m IRAC/Spitzer image of the southern
NANTEN field. The 8\,$\mu$m emission traces a sharp interface running
from north-east to south-west. It brightens at the cloud edge, showing
several elongated filaments and knots along the interface.  It also
shows a bright emission knob in the north-west, which probably
corresponds to the IRAS point source 10430-5931. The 8\,$\mu$m,
emission here is weaker than in the northern field by about a factor
5.  The IRAC emission drops rapidly towards the bulk of the cloud in
the south-east, similar to PAH emission traced by the $3.21\,\mu$m
image of \citet[][]{rathborne2002}.

In contrast to the mid-infrared emission, CO 4--3 and \CI\ integrated
intensities (Fig.\,\ref{fig-carina-s-co43-ci10}) peak south of the
interface deep inside the cloud near ($1',-8'$).  The CO 7--6 line
peaks near IRAS\,10430-5931, indicating elevated temperatures and
densities.  In contrast, the IRAS source is not prominent in \CI.

\subsubsection{Spectra along a cut through the region}

Spectra taken along a north-south cut through the southern field show
one component near $-27$\,kms$^{-1}$ with little variation in velocity
(Fig.\,\ref{fig-carina-s-co43-channels}). Near the interface,
$^{12}$CO line profiles are flat-topped indicating high optical depths
and self-absorption.
Table\,\ref{tab-carn-wco} lists the results of Gaussian fits to the
spectra at the interface position (0\farcm8,$-$7\farcm3).  The fitted
FWHMs are 4.3\,kms$^{-1}$ (CO 4--3), 3.9\,kms$^{-1}$ (CO 7--6),
3.4\,kms$^{-1}$ ($^{13}$CO 2--1), 2.1\,kms$^{-1}$ (\CI 1--0),
2.4\,kms$^{-1}$ (\CI 2--1). $^{12}$CO line widths are a factor $\sim2$
broader than the \CI\ line widths which probably reflects changes of
the optical depths.
%

Compared to the northern field, the $^{12}$CO and $^{13}$CO lines are
weaker while the \CI\ 1--0 line is stronger. The CO 4--3 peaks at only
$\sim20$\,K. The line ratios of CO 7--6/4--3 and \CI\ 2--1/1--0 at the
interface position are $0.3\pm20\%$ and $0.6\pm20$\% respectively.
These ratios are slightly lower than in the northern field, suggesting
lower temperatures and densities. However, optical depth effects and
self-absorption may also play a role.
%
%
The \CI\ 1--0/CO 4--3 ratio is 0.34 at the interface, a factor 2
higher than in the northern interface.  The large-scale survey of the
entire Carina region by \citet{zhang2001} showed variations between
0.14 and 0.45, similar to the ratios we find at angular resolutions
which are a factor $\sim5$ higher.  Gas kinetic temperatures must be
at least 30\,K to explain the $^{12}$CO 4--3 main beam temperatures.
In the southern interface, this is consistent with $T_{\rm
  ex}$(\CI)=30.5\,K ($^{36\,{\rm K}}_{26\,{\rm K}}$) derived from the
\CI\ line ratio.

The LTE analysis at the southern position indicates a C/CO abundance
ratio of 0.26, 
slightly higher than the abundance ratio in the north.
Taking into account the calibration errors, the C/CO abundance ratios
lie between $\sim0.1$ and $\sim0.4$ at both positions
(Table\,\ref{tab-carn-wco}).  Similar abundance ratios have been
observed in other Galactic star forming regions like Cepheus\,B or
NGC\,7023 \citep[see references in][]{mookerjea2006a}.

\begin{figure}[h]   
  \centering   
  \includegraphics[angle=-90,width=7cm]{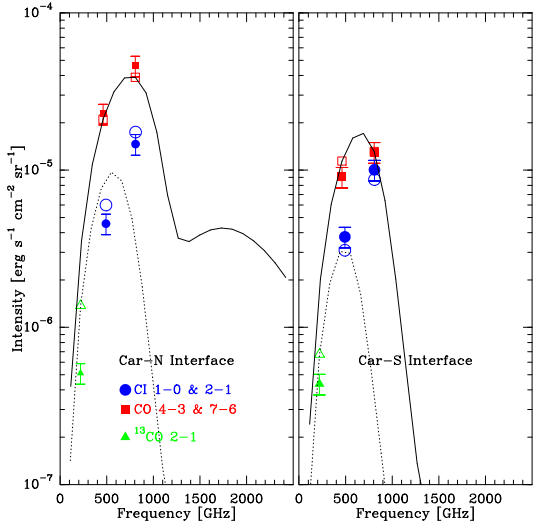}
  \caption{Integrated intensities at the two interface positions in
    Carina-North and South (Table\,\ref{tab-carn-wco}). Filled symbols
    show the observed $^{13}$CO 2--1, \CI\ 1--0, 2--1, $^{12}$CO 4--3,
    and 7--6 intensities. Error bars denote the 15\% calibration
    error.  Model results of the best fitting clumpy PDR model
    (Table\,\ref{tab-model}) are shown by solid ($^{12}$CO) and dotted
    lines ($^{13}$CO), and open symbols. }
\label{fig-cooling}   
\end{figure}   


\begin{center}   
\begin{table*}[ht]   
  \caption[]{\label{tab-model} {\small Parameters of
the clump ensembles at the two positions which fit best the observed
intensities. Columns (2) to (6) list the input parameters from which
the values given in columns (7) to (11) are derived: $\chi$ is the FUV
field in Draine units, $\langle n\rangle_{\rm ens}$ is the average
density of the clump ensemble, $M_{\rm cl}^{\rm min}$, $M_{\rm
cl}^{\rm max}$, $M_{\rm tot}$ are the minimum and maximum clump mass,
and the total ensemble mass, respectively. $n_{\rm{0,min}}$ and
$n_{\rm{0,max}}$ are the smallest and largest clump surface densities,
$R_{\rm min}$ and $R_{\rm max}$ are the smallest and largest clump
radii, and $\phi_A$ is the beam-filling factor of the clump
ensemble. }}
\begin{tabular}{lllllllllrrrrrr}   
\hline \hline   
  $\Delta\alpha/\Delta\delta$ 
  & $\chi$ & $\langle n\rangle_{\rm ens}$ 
  & $M_{\rm cl}^{\rm min}$ & $M_{\rm cl}^{\rm max}$ & $M_{\rm tot}$ 
  & $n_{\rm{0,min}}$   & $n_{\rm{0,max}}$ & $R_{\rm min}$ & $R_{\rm max}$
  & $\phi_A$ \\
  $(',')$      
  &        & cm$^{-3}$ 
  & \msol                  & \msol                  & \msol         
  & cm$^{-3}$          & cm$^{-3}$        & pc & pc 
  & \\
  (1) & (2) & (3) & (4) & (5) & (6) & (7) & (8) & (9) & (10) & (11) \\
\hline
  $-11.7/6.8$  & $10^{3.5}$ & $2\,10^5$ 
  & $10^{-2}$ & $10^2$ & 400 
  & $5.3\,10^4$ & $8.8\,10^5$   &   $3.7\,10^{-3}$ & $2.1\,10^{-1}$ 
  & 3.9 \\ 
  $+0.8/-7.2$  & $10^{2.5}$ & $2\,10^5$ 
  & $10^{-2}$ & $10^2$ & 220
  & $5.3\,10^4$ & $8.8\,10^5$ & $3.7\,10^{-3}$ & 0.2  
  & 2.1 \\
\noalign{\smallskip} \hline \noalign{\smallskip}   
\end{tabular}   
\end{table*}   
\end{center}   

\begin{center}   
\begin{table*}[ht]   
  \caption[]{\label{tab-int-ratios} 
    {\small Ratios of observed integrated intensities
over integrated intensities predicted by the clumpy PDR model.  The
input parameters used are given in Table\,\ref{tab-model}; deviations
from this setting are listed in column (7).  }}
\begin{tabular}{lllllllllrrrrr}   
\hline \hline   
  $\Delta\alpha/\Delta\delta$ & $^{13}$CO & \multicolumn{2}{c}{CO} & \multicolumn{2}{c}{\CI} \\
  $(',')$ & 2--1 & 4--3 & 7--6 & 1--0 & 2--1 \\
  (1) & (2) & (3) & (4) & (5) & (6) & (7) \\
\hline
 $-11.7/6.8$ & 0.4 & 1.1 & 1.2 & 0.8 & 0.8 & \\ 
 $-11.7/6.8$ & 0.3 & 0.9 & 0.8 & 0.7 & 0.8 & $M_{\rm min}=10\,^{-3}$\,\msun \\
 $-11.7/6.8$ & 0.4 & 1.3 & 1.7 & 0.8 & 1   & $M_{\rm min}=10\,^{-1}$\,\msun \\
 $-11.7/6.8$ & 0.5 & 1.7 & 1.5 & 1.4 & 1.5 & $M_{\rm max}=10^3$\,\msun \\
 $-11.7/6.8$ & 0.3 & 0.8 & 1 & 0.5 & 0.5 & $M_{\rm max}=10^1$\,\msun \\
%
\hline
 $+0.8/-7.2$ & 0.6 & 0.8 & 1 & 1.2 & 1.2 \\
\noalign{\smallskip} \hline \noalign{\smallskip}   
\end{tabular}   
\end{table*}   
\end{center}

\section{Clumpy interface regions} 

Figure\,\ref{fig-cooling} shows the cooling intensities\footnote{We
  used $\int I_\nu d\nu$
  [erg\,s$^{-1}$\,cm$^{-2}$sr$^{-1}$]=$\frac{2{\rm k}\nu^3}{c^3}\int T
  d$v [K\,kms$^{-1}$].} observed at the two interface positions
(Table\,\ref{tab-carn-wco}) along the cuts presented in the previous
section. At both positions, the 7-6 intensity exceeds
  the 4-3 intensity. As this transition has a critical density of
  $1.4\,10^6$\,cm$^{-3}$ and an upper-level energy corresponding to
  157\,K, temperatures and densities are expected to be similarly
  high.
%
%

In the following, we compare the observed emission with the
predictions of a clumpy PDR model \citep{cubick2005,cubick2007}.  We
assume that the emission stems from an ensemble of spherically
symmetric clumps within the beam.  The clumps follow the canonical
clump mass and mass size distributions: $dN/dM\propto M^{-\alpha}$ and
$M\propto R^{\gamma}$. For the clump mass spectrum
  derived from molecular line observations, $\alpha\sim1.8$ in a large
  number of Galactic clouds \citep[e.g.][]{kramer1998,simon2001}.
%
For the power law describing the mass-size distribution,
$\gamma\sim2.3$ in several Galactic clouds
\citep[e.g.][]{heithausen1998}.
As a result, the clump density distribution is $n\propto
M^{1-3/\gamma}$,
i.e. the density increases with decreasing clump mass.
%
For the nearby clouds L\,1457 and the Polaris Flare
  presented in the Kramer and Heithausen papers, the canonical clump
  mass and mass size distributions continue down to masses of less
  than $10^{-3}$\,\msun\ and radii of less than $10^{-2}$\,pc.

%
%
%

  The emission from a clump of given mass and density is calculated by
  the KOSMA-$\tau$ model \citep{roellig2006,stoerzer1996}.  The radial
  density structure $n(r)$ of a model clump with radius $R_0$ is given
  by $n(r)\propto n_0\,(r/R_0)^{-1.5}$ for $0.2\le r/R_0\le 1$ and
  $n(r)=11.18\,n_0=$\,const. for $r/R_0\le0.2$. The clump averaged
  density $\langle n\rangle$ is about twice the density at the clump
  surface, called henceforth {\it surface density} $n_0$: $\langle
  n\rangle \sim 1.91\,n_0$.  PDR models of individual clumps were
  calculated on a grid of FUV fields $\log(\chi)=0, 0.5,$ to 6, clump
  surface densities $\log(n_0/{\rm cm}^{-3})=2, 2.5,$ to 7, and clump
  masses $\log(M_{\rm cl}/$\msun$)=-3, -2.5,$ to 2.
The parameters controlling the ensemble model are the
  FUV field, the average ensemble density, the total ensemble mass,
  and the lower and upper clump mass limits.
Here, we assume that all clumps are heated by a constant FUV field, as
derived from the HIRES observations presented above.
We chose a lower mass cutoff of the ensemble of
  $10^{-2}$\,\msun\ and an upper cutoff of 100\,\msun. Below, we show
  that the exact choice of these limits does however not have a strong
  impact on the resulting model intensities \citep[cf.][]{cubick2007}.
  To fit the observed intensity ratios \CI\ 2--1/1--0 and CO
  7--6/4--3, the only free parameter is then the average density of
  the clump ensemble $\langle n\rangle_{\rm ens}$.

We adapted the total ensemble mass to fit the absolute intensities.
The best fitting models and their parameters are shown in
Figure\,\ref{fig-cooling} and in Table\,\ref{tab-model}. Table
\ref{tab-int-ratios} lists the ratios of observed over modelled
intensities for all 5 observed transitions. For a perfectly fitting
model, they would all equal $1\pm15\%$, reflecting the estimated
calibration error of the observed data.
%

\paragraph{Northern interface.}

An average ensemble density of $2\,10^5$\,cm$^{-3}$ allows us to
reproduce the observed $^{12}$CO 7--6/4--3 and \CI\ 2--1/1--0 line
ratios very well at the northern interface position. Lower densities
do not reproduce the high CO 7--6/4--3 line ratios, but rather lead to
a peak of the cooling curve at $J\le5$.  The absolute intensities of
the $^{12}$CO and $^{12}$C lines can be reproduced to within 20\% for
an ensemble mass of 400\,\msol. The strongest deviation occurs for
$^{13}$CO; the observed $^{13}$CO 2--1 line intensity is only 40\% of
the modelled intensity.

%
%
%

The geometrical beam-filling factor of the clump ensemble is defined
as the ratio of the sum of all clump areas over the beam area, i.e.
$\phi_A=\Omega_{\rm ens}/\Omega_{\rm beam}$, with $\Omega_{\rm
  ens}=\sum \pi R_{\rm cl}^2/d^2$ at distance $d$. The beam-filling
factor is 3.9 at the northern interface position
(Table\,\ref{tab-model}).
We assume that clumps don't overlap, i.e. that their
  emission escapes without being absorbed by foreground clumps, and
  there is no interclump medium. Clumps have a constant velocity FWHM
  of $\Delta{\rm v}=1.7$\,kms$^{-1}$.
  The clumpy ensemble models developed by \citet{cubick2007} do not
  yet predict line profiles.  Table\,\ref{tab-model} gives the
  densities and radii of the smallest and largest clumps of the
  ensemble. The mass of the smallest clump is set to $10^{-2}$\,\msun.
  The virial mass of the smallest clump, $M_{\rm vir}\propto
  R\Delta{\rm v}^2$ is a factor 400 larger. In other words, the
  velocity width of these clumps are far too large for pure
  gravitational virialization. The external pressure which would be
  needed to confine these clumps, $10^8$\,Kcm$^{-3}$, is far too large
  to be maintained by an interclump medium. The large ratio indicates
  instead that these clumps, if they do exist, are transient features
  of the turbulent gas and evaporating on very short time scales:
  $t_{\rm evap}\approx R/\sigma_{\rm v} = R/{\rm FWHM}
  \sqrt{8\ln2}\approx 5000\,$yrs.  

The modelled intensities depend only weakly on the
  lower and upper mass limits of the ensemble as shown in
  Table\,\ref{tab-int-ratios}.  Varying one of the mass limits by one
  order of magnitude while keeping all other 4 input parameters listed
  in Table\,\ref{tab-model} unchanged, the absolute intensities of the
  5 transitions vary by less than a factor of $\sim2$ relative to the
  best fitting solution. This is discussed in more detail in
  \citet{cubick2007}.

The run of $^{12}$CO intensities with rotational
  number $J$, i.e. the CO cooling curve, predicted by the model
  (Fig.\,\ref{fig-cooling}) shows a maximum at $J=7$ and a steep drop
  at higher rotational numbers up to $J=11$ by more than a magnitude.
  A secondary peak of the cooling curve is seen at $J=15$; it is
  almost a factor 10 weaker than the first maximum. The second peak
  shows up only for FUV fields above $\sim10^3$
  and it stems predominantly from the surface regions of the small,
  dense clumps.  These clumps are large in number, have a large total
  surface area, which is heated to high temperatures, sufficient to
  excite these high-$J$ lines.
  The abundance of CO is increased in the hot gas of the \HI/H$_2$
  surface layer, near the OH density peak, due to secondary production
  paths which become important at high temperatures
  \citep{sternberg1995}:
    \begin{eqnarray}
      {\rm OH}+{\rm C}^+ & \rightarrow & {\rm CO}+{\rm H}^+ 
           \hspace*{0.5cm} {\rm (R13)} \nonumber \\
      {\rm OH}+{\rm C}^+ & \rightarrow & {\rm CO}^+ +{\rm H}
                           \rightarrow {\rm CO}+{\rm H}^+
            \hspace*{0.5cm} {\rm (R125,R126)} \nonumber\\
      {\rm CO}^+ +{\rm H}_2 & \rightarrow & {\rm HCO}^+ + {\rm H} 
           \hspace*{0.5cm} {\rm (R127)} \nonumber \\
      {\rm HCO}^+ + e & \rightarrow & {\rm CO} + {\rm H}      
           \hspace*{0.5cm} {\rm (R129)} \nonumber
    \end{eqnarray} 
We suspect that these reactions lead to the secondary peak of the
cooling curve shown in Figure\,\ref{fig-cooling}.
%

%
%


\paragraph{Southern interface.} At the southern
  interface position, the FUV field is only $10^{2.5}\,\chi_0$. An
  average ensemble density of $2\,10^5$\,cm$^{-3}$ reproduces the
  observed $^{12}$CO 7--6/4--3 and \CI\ 2--1/1--0 line ratios. Within
  40\%, the clumpy model is also consistent with the absolute
  intensities of all 5 tracers for a total ensemble mass of
  220\,\msol\ (Fig.\,\ref{fig-cooling},
  Tables\,\ref{tab-int-ratios},\ref{tab-model}). Due to the weaker FUV
  field compared to the northern position, the CO cooling curve peaks
  at slightly lower frequencies and a second peak is not visible.
  Similar to the northern position, the strongest deviation occurs for
  the $^{13}$CO 2--1 lines. 





\section{Summary} 

We have mapped the emission of atomic carbon and CO in two fields of
the GMCs surrounding $\eta$Car. Combining CO 4--3 and \CI\ 1--0 data
with CO 7--6 and \CI\ 2--1 data allows us to study the excitation
conditions of both gas tracers.  The northern $4'\times4'$ field lies
adjacent to the compact OB cluster Tr\,14 and is associated with the
\HII\ regions Car\,I-E and Car\,I-W. The spectral lines show a rich
kinematical structure, with several velocity components along the
lines of sight spread between $-10$ and $-30\,$kms$^{-1}$.  The
southern $4'\times4'$ field lies on a molecular ridge south of
$\eta$Car. Here, spectral lines show only one rather narrow Gaussian
shaped velocity component. The FUV field derived from HIRES/IRAS
far-infrared data is $\chi\sim10^{2.5}$, about a factor 10 weaker than
in the northern field. IRAC/Spitzer maps at 8$\mu$m show the detailed
structure of the FUV illuminated dust. The edge of the southern region
is clearly delineated by the $8\mu$m emission, while the northern
region shows a more complex spatial structure.

In both regions, CO and \CI\ emission is co-extensive. Both species
appear to trace the bulk of molecular gas, rather than the interface
regions. This coincidence has been found in many of the previous
large-scale studies of Galactic clouds conducted with the Mt.Fuji,
KOSMA, and other telescopes. \citet{papadopoulos2004} stress the
importance of cosmic rays in raising the C/CO abundance ratio and
argue that dynamic and non-equilibrium chemistry processes explain the
ubiquity of \CI\ within molecular clouds as well as at PDR interfaces.
Here, we argue that steady state, clumpy PDR models provide an
alternative explanation.

The northern and southern Carina regions both show a slight
brightening of the CO 7--6 and \CI\ 2--1 transitions relative to the
4--3 and 1--0 transitions near the interfaces. In the south-east part
of the northern region, the upper transitions brighten towards the
interface (Fig.\,\ref{fig-carina-n-co43-ci10}).  These transitions
also show increased intensities near the embedded IRAS point source in
the southern region (Fig.\,\ref{fig-carina-s-co43-ci10}).

We selected two cuts from the cloud cores through the interface
regions, roughly pointing towards the illuminating sources.  The
southern cut lies to the east of IRAS\,10430-5931, avoiding its
complicated structure and additional heating sources. The northern cut
runs from the peak of \CII\ emission \citep{brooks2003} through the
\HII\ regions Car\,I-W and I-E across the edge-on interface towards
the center of the Tr\,14 cluster.

The peak $^{12}$CO line temperatures along the northern cut indicate
kinetic gas temperatures of at least 50\,K, while a lower limit of
30\,K is found along the southern cut.

We selected two positions at the northern and southern interfaces for
a detailed analysis.  The $^{12}$CO 7--6/4--3 and \CI\ 2--1/1--0
intensity ratios at these positions are 0.3--0.4 and 0.6--0.7,
respectively.  We use PDR models to interpret the observed line
intensities. Our models assume that all observed emission stems from
an ensemble of spherically symmetric PDR clumps and there is no
emission from an interclump medium or a diffuse halo surrounding the
clouds. PDR clumps are modelled using the stationary KOSMA-$\tau$ code
\citep{roellig2006}.  The clump distributions follow the canonical
mass and radius distributions found for molecular clouds. At the two
positions analyzed in detail, we find that clumpy PDR models are
consistent with the observed absolute intensities of the $^{12}$CO and
\CI\ lines to within 20\%, i.e. at about the calibration accuracy.

Since the line ratios observed at the two interface positions are
fairly typical for the entire observed regions, we conclude that
stationary, clumpy PDR models can simultaneously reproduce the
observed \CI\ and CO emission of the lower transitions and the upper
CO 7--6 and \CI\ 2--1 transitions.

However, the observed $^{13}$CO 2--1 intensities are
  only about half of the modelled intensities at the northern and
  southern positions, respectively. Interestingly, \citet{pineda2007}
  report a similar finding. They analyzed $^{12}$CO and \CI\ emission
  observed in the dark cloud globule B\,68. This emission can be
  reproduced using a single KOSMA-$\tau$ spherically symmetric PDR
  model, while the observed $^{13}$CO intensities are a factor $\sim2$
  too low.

To study the gas under a broad range of conditions, we plan to extend
the present \CI\ and CO maps to fully cover the GMCs surrounding
$\eta$Car using the SMART 492/810\,GHz multipixel array receiver at
NANTEN2 \citep{graf2003}.  Future observations with APEX and Herschel
will tell in how far the emission of other species tracing the
detailed photo-chemical network is also consistent with stationary,
clumpy PDR models.

\begin{acknowledgements} 
  
  We thank Kate Brooks and Nicola Schneider for making their partly
  unpublished SEST $^{13}$CO 2--1 data available to us. We also would
  like to thank an anonymous referee for insightful comments which
  helped to improve on our arguments. We made use of the
  NASA/IPAC/IRAS/HiRES data reduction facilities.  Data reduction of
  the spectral line data was done with the {\tt gildas} software
  package supported at IRAM (see {\tt
    http://www.iram.fr/IRAMFR/GILDAS}). This research has made use of
  NASA's Astrophysics Data System Abstract Service.

  This work is financially supported in part by a Grant-in-Aid for
  Scientific Research from the Ministry of Education, Culture, Sports,
  Science and Technology of Japan (No.¥ 15071203) and from JSPS (No.¥
  14102003 and No.¥ 18684003), and by the JSPS core-to-core program
  (No.¥ 17004).
  This work is also financially supported in part by the grant
  SFB\,494 of the Deutsche Forschungsgemeinschaft, the Ministerium
  f\"ur Innovation, Wissenschaft, Forschung und Technologie des Landes
  Nordrhein-Westfalen and through special grants of the Universit\"at
  zu K\"oln and Universit\"at Bonn.
  L. B. and J. M. acknowledge support from the Chilean Center for
  Astrophysics FONDAP 15010003.

\end{acknowledgements}   
   
\bibliographystyle{aa} 
\bibliography{aamnem99,7815} 

\begin{thebibliography}{63}
\expandafter\ifx\csname natexlab\endcsname\relax\def\natexlab#1{#1}\fi

\bibitem[{Aumann {et~al.}(1990)Aumann, Fowler, \& Melnyk}]{aumann1990}
Aumann, H., Fowler, J., \& Melnyk, M. 1990, AJ, 99, 1674

\bibitem[{{Bayet} {et~al.}(2006){Bayet}, {Gerin}, {Phillips}, \&
  {Contursi}}]{bayet2006}
{Bayet}, E., {Gerin}, M., {Phillips}, T.~G., \& {Contursi}, A. 2006, \aap, 460,
  467

\bibitem[{{Bensch}(2006)}]{bensch2006}
{Bensch}, F. 2006, \aap, 448, 1043

\bibitem[{Bensch {et~al.}(2003)Bensch, Leuenhagen, Stutzki, \&
  Schieder}]{bensch2003}
Bensch, F., Leuenhagen, U., Stutzki, J., \& Schieder, R. 2003, A\&A, 591, 1013

\bibitem[{{Bensch} {et~al.}(2001){Bensch}, {Stutzki}, \&
  {Heithausen}}]{bensch2001aa365-285}
{Bensch}, F., {Stutzki}, J., \& {Heithausen}, A. 2001, \aap, 365, 285

\bibitem[{Bohlin {et~al.}(1978)Bohlin, Savage, \& Drake}]{bohlin1978}
Bohlin, R.~C., Savage, B.~D., \& Drake, J.~F. 1978, ApJ, 224, 132

\bibitem[{Boiss\'{e}(1990)}]{boisse1990}
Boiss\'{e}, P. 1990, A\&A, 228, 483

\bibitem[{Brooks {et~al.}(2003)Brooks, Cox, Schneider, Storey, Poglitsch,
  Geiss, \& Bronfman}]{brooks2003}
Brooks, K., Cox, P., Schneider, N., {et~al.} 2003, A\&A, 412, 751

\bibitem[{{Brooks} {et~al.}(2001){Brooks}, {Storey}, \&
  {Whiteoak}}]{brooks2001}
{Brooks}, K.~J., {Storey}, J.~W.~V., \& {Whiteoak}, J.~B. 2001, \mnras, 327, 46

\bibitem[{{Cox}(1995)}]{cox1995}
{Cox}, P. 1995, in Revista Mexicana de Astronomia y Astrofisica Conference
  Series, ed. V.~{Niemela}, N.~{Morrell}, \& A.~{Feinstein}, 105

\bibitem[{Cubick(2005)}]{cubick2005}
Cubick, M. 2005, Modelling the FIR emission of the Milky Way, Diploma thesis
  (Universit{\"{a}}t zu K{\"{o}}ln)

\bibitem[{Cubick {et~al.}(2007)Cubick, Stutzki, Ossenkopf, Kramer, \&
  R\"ollig}]{cubick2007}
Cubick, M., Stutzki, J., Ossenkopf, V., Kramer, C., \& R\"ollig, M. 2007, A\&A,
  in prep.

\bibitem[{{Dale} {et~al.}(2001){Dale}, {Helou}, {Contursi}, {Silbermann}, \&
  {Kolhatkar}}]{dale2001}
{Dale}, D.~A., {Helou}, G., {Contursi}, A., {Silbermann}, N.~A., \&
  {Kolhatkar}, S. 2001, apj, 549, 215

\bibitem[{Draine \& Bertoldi(1996)}]{draine1996}
Draine, B. \& Bertoldi, F. 1996, ApJ, 468, 269

\bibitem[{{Draine}(1978)}]{draine1978}
{Draine}, B.~T. 1978, \apjs, 36, 595

\bibitem[{Fixsen {et~al.}(1999)Fixsen, Bennett, \& Mather}]{fixsen1999}
Fixsen, D.~J., Bennett, C.~L., \& Mather, J.~C. 1999, ApJ, 526, 207

\bibitem[{Frerking {et~al.}(1982)Frerking, Langer, \& Wilson}]{frerking1982}
Frerking, M.~A., Langer, W.~D., \& Wilson, R.~W. 1982, ApJ, 262, 590

\bibitem[{{Grabelsky} {et~al.}(1988){Grabelsky}, {Cohen}, {Bronfman}, \&
  {Thaddeus}}]{grabelsky1988}
{Grabelsky}, D.~A., {Cohen}, R.~S., {Bronfman}, L., \& {Thaddeus}, P. 1988,
  \apj, 331, 181

\bibitem[{{Graf} {et~al.}(2003){Graf}, {Heyminck}, {Michael}, {Stanko},
  {Honingh}, {Jacobs}, {Schieder}, {Stutzki}, \& {Vowinkel}}]{graf2003}
{Graf}, U., {Heyminck}, S., {Michael}, E., {et~al.} 2003, in {Millimeter and
  Submillimeter Detectors for Astronomy. Edited by Phillips, Thomas G.;
  Zmuidzinas, Jonas. Proceedings of the SPIE}, Vol. 4855, 322

\bibitem[{{Greve} {et~al.}(1998){Greve}, {Kramer}, \& {Wild}}]{greve1998}
{Greve}, A., {Kramer}, C., \& {Wild}, W. 1998, \aaps, 133, 271

\bibitem[{Griffin {et~al.}(1986)Griffin, Ade, Orton, Robson, Gear, Nolt, \&
  Radostitz}]{griffin1986}
Griffin, M.~J., Ade, P., Orton, G., {et~al.} 1986, Icarus, 65, 244

\bibitem[{{Hegmann} {et~al.}(2007){Hegmann}, {Kegel}, \&
  {Sedlmayr}}]{hegmann2007}
{Hegmann}, M., {Kegel}, W.~H., \& {Sedlmayr}, E. 2007, \aap, 469, 223

\bibitem[{Heithausen {et~al.}(1998)Heithausen, Bensch, Stutzki, Falgarone, \&
  Panis}]{heithausen1998}
Heithausen, A., Bensch, F., Stutzki, J., Falgarone, E., \& Panis, J. 1998,
  A\&A, 331, 65

\bibitem[{Helou {et~al.}(1988)Helou, Khan, Malek, \& Boehmer}]{helou1988}
Helou, G., Khan, I., Malek, L., \& Boehmer, L. 1988, apj, 68, 151

\bibitem[{{Hollenbach} \& {Tielens}(1999)}]{hollenbach1999}
{Hollenbach}, D.~J. \& {Tielens}, A.~G.~G.~M. 1999, Reviews of Modern Physics,
  71, 173

\bibitem[{{Jakob} {et~al.}(2007){Jakob}, {Kramer}, {Simon}, {Schneider},
  {Ossenkopf}, {Bontemps}, {Graf}, \& {Stutzki}}]{jakob2007}
{Jakob}, H., {Kramer}, C., {Simon}, R., {et~al.} 2007, \aap, 461, 999

\bibitem[{{Juvela} {et~al.}(2001){Juvela}, {Padoan}, \&
  {Nordlund}}]{juvela2001}
{Juvela}, M., {Padoan}, P., \& {Nordlund}, {\AA}. 2001, \apj, 563, 853

\bibitem[{{Kramer} {et~al.}(2005){Kramer}, {Mookerjea}, {Bayet},
  {Garcia-Burillo}, {Gerin}, {Israel}, {Stutzki}, \& {Wouterloot}}]{kramer2005}
{Kramer}, C., {Mookerjea}, B., {Bayet}, E., {et~al.} 2005, \aap, 441, 961

\bibitem[{{Kramer} {et~al.}(1998){Kramer}, {Stutzki}, {R{\"o}hrig}, \&
  {Corneliussen}}]{kramer1998}
{Kramer}, C., {Stutzki}, J., {R{\"o}hrig}, R., \& {Corneliussen}, U. 1998,
  \aap, 329, 249

\bibitem[{{Mangum}(1993)}]{mangum1993}
{Mangum}, J.~G. 1993, \pasp, 105, 117

\bibitem[{{Meixner} \& {Tielens}(1993)}]{meixner1993}
{Meixner}, M. \& {Tielens}, A.~G.~G.~M. 1993, \apj, 405, 216

\bibitem[{{Milam} {et~al.}(2005){Milam}, {Savage}, {Brewster}, {Ziurys}, \&
  {Wyckoff}}]{milam2005}
{Milam}, S.~N., {Savage}, C., {Brewster}, M.~A., {Ziurys}, L.~M., \& {Wyckoff},
  S. 2005, \apj, 634, 1126

\bibitem[{{Mizutani} {et~al.}(2002){Mizutani}, {Onaka}, \&
  {Shibai}}]{mizutani2002}
{Mizutani}, M., {Onaka}, T., \& {Shibai}, H. 2002, \aap, 382, 610

\bibitem[{{Mizutani} {et~al.}(2004){Mizutani}, {Onaka}, \&
  {Shibai}}]{mizutani2004}
{Mizutani}, M., {Onaka}, T., \& {Shibai}, H. 2004, \aap, 423, 579

\bibitem[{{Mookerjea} {et~al.}(2006){Mookerjea}, {Kramer}, {R{\"o}llig}, \&
  {Masur}}]{mookerjea2006a}
{Mookerjea}, B., {Kramer}, C., {R{\"o}llig}, M., \& {Masur}, M. 2006, \aap,
  456, 235

\bibitem[{{Nakagawa} {et~al.}(1998){Nakagawa}, {Yui}, {Doi}, {Okuda}, {Shibai},
  {Mochizuki}, {Nishimura}, \& {Low}}]{nakagawa1998}
{Nakagawa}, T., {Yui}, Y.~Y., {Doi}, Y., {et~al.} 1998, \apjs, 115, 259

\bibitem[{{Oberst} {et~al.}(2006){Oberst}, {Parshley}, {Stacey}, {Nikola},
  {L{\"o}hr}, {Harnett}, {Tothill}, {Lane}, {Stark}, \& {Tucker}}]{oberst2006}
{Oberst}, T.~E., {Parshley}, S.~C., {Stacey}, G.~J., {et~al.} 2006, \apjl, 652,
  L125

\bibitem[{{Oka} {et~al.}(2004){Oka}, {Iwata}, {Maezawa}, {Ikeda}, {Ito},
  {Kamegai}, {Sakai}, \& {Yamamoto}}]{oka2004}
{Oka}, T., {Iwata}, M., {Maezawa}, H., {et~al.} 2004, apj, 602, 803

\bibitem[{{Papadopoulos} {et~al.}(2004){Papadopoulos}, {Thi}, \&
  {Viti}}]{papadopoulos2004}
{Papadopoulos}, P.~P., {Thi}, W.-F., \& {Viti}, S. 2004, MNRAS, 351, 147

\bibitem[{Pardo {et~al.}(2001)Pardo, Cernicharo, \& Serabyn}]{pardo2001}
Pardo, J., Cernicharo, J., \& Serabyn, E. 2001, IEEE Transactions on Antennas
  and Propagation, 49, 1683

\bibitem[{{Pardo} {et~al.}(2005){Pardo}, {Serabyn}, \& {Wiedner}}]{pardo2005}
{Pardo}, J.~R., {Serabyn}, E., \& {Wiedner}, M.~C. 2005, Icarus, 178, 19

\bibitem[{{Pineda} \& {Bensch}(2007)}]{pineda2007}
{Pineda}, J.~L. \& {Bensch}, F. 2007, \aap, 470, 615

\bibitem[{{Rathborne} {et~al.}(2002){Rathborne}, {Burton}, {Brooks}, {Cohen},
  {Ashley}, \& {Storey}}]{rathborne2002}
{Rathborne}, J.~M., {Burton}, M.~G., {Brooks}, K.~J., {et~al.} 2002, \mnras,
  331, 85

\bibitem[{{R{\"o}llig} {et~al.}(2007){R{\"o}llig}, {Abel}, {Bell}, {Bensch},
  {Black}, {Ferland}, {Jonkheid}, {Kamp}, {Kaufman}, {Le Bourlot}, {Le Petit},
  {Meijerink}, {Morata}, {Ossenkopf}, {Roueff}, {Shaw}, {Spaans}, {Sternberg},
  {Stutzki}, {Thi}, {van Dishoeck}, {van Hoof}, {Viti}, \&
  {Wolfire}}]{roellig2007}
{R{\"o}llig}, M., {Abel}, N.~P., {Bell}, T., {et~al.} 2007, \aap, 467, 187

\bibitem[{{R{\"o}llig} {et~al.}(2006){R{\"o}llig}, {Ossenkopf}, {Jeyakumar},
  {Stutzki}, \& {Sternberg}}]{roellig2006}
{R{\"o}llig}, M., {Ossenkopf}, V., {Jeyakumar}, S., {Stutzki}, J., \&
  {Sternberg}, A. 2006, \aap, 451, 917

\bibitem[{{Sakai} {et~al.}(2006){Sakai}, {Oka}, \& {Yamamoto}}]{sakai2006}
{Sakai}, T., {Oka}, T., \& {Yamamoto}, S. 2006, \apj, 649, 268

\bibitem[{{Schulz} {et~al.}(2007){Schulz}, {Henkel}, {Muders}, {Mao},
  {R{\"o}llig}, \& {Mauersberger}}]{schulz2007}
{Schulz}, A., {Henkel}, C., {Muders}, D., {et~al.} 2007, \aap, 466, 467

\bibitem[{Simon {et~al.}(2007)Simon, Graf, Kramer, Stutzki, \&
  Onishi}]{simon2007}
Simon, R., Graf, U., Kramer, C., Stutzki, J., \& Onishi, T. 2007, in NANTEN
  technical report, 13.2.2007, Vol.~1, 1

\bibitem[{{Simon} {et~al.}(2001){Simon}, {Jackson}, {Clemens}, {Bania}, \&
  {Heyer}}]{simon2001}
{Simon}, R., {Jackson}, J.~M., {Clemens}, D.~P., {Bania}, T.~M., \& {Heyer},
  M.~H. 2001, \apj, 551, 747

\bibitem[{{Smith}(2006{\natexlab{a}})}]{smith2006}
{Smith}, N. 2006{\natexlab{a}}, \mnras, 367, 763

\bibitem[{{Smith}(2006{\natexlab{b}})}]{smith2006b}
{Smith}, N. 2006{\natexlab{b}}, \apj, 644, 1151

\bibitem[{{Smith} \& {Brooks}(2007)}]{smith_brooks2007}
{Smith}, N. \& {Brooks}, K.~J. 2007, \mnras, 379, 1279

\bibitem[{{Smith} {et~al.}(2005){Smith}, {Churchwell}, {Whitney}, {Meade},
  {Babler}, {Bally}, {Stassun}, {Morse}, \& {Gehrz}}]{smith2005aas}
{Smith}, N., {Churchwell}, E.~B., {Whitney}, B., {et~al.} 2005, in Bulletin of
  the American Astronomical Society, 439

\bibitem[{{Spaans}(1996)}]{spaans1996}
{Spaans}, M. 1996, \aap, 307, 271

\bibitem[{{Spaans} \& {van Dishoeck}(1997)}]{spaans1997}
{Spaans}, M. \& {van Dishoeck}, E.~F. 1997, \aap, 323, 953

\bibitem[{Sternberg \& Dalgarno(1995)}]{sternberg1995}
Sternberg, A. \& Dalgarno, A. 1995, ApJ, 99, 565

\bibitem[{St{\"o}rzer {et~al.}(1996)St{\"o}rzer, Stutzki, \&
  Sternberg}]{stoerzer1996}
St{\"o}rzer, H., Stutzki, J., \& Sternberg, A. 1996, A\&A, 310, 592

\bibitem[{St{\"o}rzer {et~al.}(1997)St{\"o}rzer, Stutzki, \&
  Sternberg}]{stoerzer1997}
St{\"o}rzer, H., Stutzki, J., \& Sternberg, A. 1997, A\&A, 323, 13

\bibitem[{Tatematsu {et~al.}(1999)Tatematsu, Jaffe, Plume, \&
  Evans}]{tatematsu1999}
Tatematsu, K., Jaffe, D., Plume, R., \& Evans, N. 1999, ApJ, 526, 295

\bibitem[{Tielens \& Hollenbach(1985)}]{tielens1985}
Tielens, A. \& Hollenbach, D. 1985, ApJ, 291, 722

\bibitem[{Weiss {et~al.}(2003)Weiss, Henkel, Downes, \& Walter}]{weiss2003}
Weiss, A., Henkel, C., Downes, D., \& Walter, F. 2003, A\&A, 409, 41

\bibitem[{{Yonekura} {et~al.}(2005){Yonekura}, {Asayama}, {Kimura}, {Ogawa},
  {Kanai}, {Yamaguchi}, {Barnes}, \& {Fukui}}]{yonekura2005}
{Yonekura}, Y., {Asayama}, S., {Kimura}, K., {et~al.} 2005, \apj, 634, 476

\bibitem[{Zhang {et~al.}(2001)Zhang, Lee, Bolatto, \& Stark}]{zhang2001}
Zhang, X., Lee, Y., Bolatto, A., \& Stark, A.~A. 2001, ApJ, 553, 274

\end{thebibliography}

\end{document}